\documentclass[10pt,letterpaper,twocolumn]{article}
\usepackage[margin=0.75in,columnsep=0.25in]{geometry}
\usepackage[T1]{fontenc}
\usepackage{lmodern}
\usepackage{amsmath,amssymb}
\usepackage{graphicx}
\usepackage{booktabs}
\usepackage[labelfont=bf,labelsep=period,font=small]{caption}
\usepackage{microtype}
\usepackage{authblk}
\usepackage[numbers,sort&compress]{natbib}
\usepackage[colorlinks=true,linkcolor=black,citecolor=black,urlcolor=blue]{hyperref}

\newcommand{\Meff}{4\pi M_{\mathrm{eff}}}
\newcommand{\Ms}{M_{s}}
\newcommand{\Keff}{K_{\mathrm{eff}}}
\newcommand{\vect}[1]{\mathbf{#1}}
\newcommand{\Nia}{Ni(300\,\AA)}
\newcommand{\Nib}{Ni(1000\,\AA)}

\title{\textbf{Perpendicular Anisotropy and \texorpdfstring{$g$}{g}-Factor of Sputtered Ni Films on Si(111) Studied by Angle-Dependent Ferromagnetic Resonance}}

\author[1]{Dushyant Singh}
\author[1]{Tharundev V V}
\author[1]{Subha Maity}
\author[1]{Subhavenkadesh SS}
\author[1]{Vigneshwari SR}
\author[1,*]{Krista R. Khiangte}

\affil[1]{Department of Physics, Indian Institute of Technology Gandhinagar, Gandhinagar, Gujarat 382055, India}
\affil[*]{Corresponding author: krista.khiangte@iitgn.ac.in}
\date{}

\begin{document}
\twocolumn[
\begin{@twocolumnfalse}
\maketitle

\begin{abstract}
\noindent Ni films with thicknesses of 312 and 1011\,\AA{} were grown on Si(111) by dc magnetron sputtering at room temperature and studied by atomic force microscopy, grazing-incidence X-ray diffraction, X-ray reflectivity, magnetometry and angle-dependent ferromagnetic resonance. Both films are polycrystalline fcc Ni with smooth surfaces, a bulk-like (111) lattice spacing and a saturation magnetization of ${\approx}480$\,emu\,cm$^{-3}$, and both are magnetized in plane. The angular dependence of the resonance field, from the film plane to the film normal, is described by a uniaxial anisotropy model with first- and second-order terms. From this model, $g = 2.160\pm0.005$ and $2.101\pm0.014$ and effective perpendicular anisotropies of $(6.0\pm0.4)\times10^{5}$ and $(4.5\pm0.6)\times10^{5}$\,erg\,cm$^{-3}$ were obtained for the thinner and thicker film, respectively. The conventional analysis of only the in-plane and perpendicular resonances gives $g$-factors that are lower by 0.02--0.04. The perpendicular linewidths give upper limits for the Gilbert damping of $\alpha \le 0.030$ and $\le 0.038$.

\end{abstract}
\vspace{1.5em}
\end{@twocolumnfalse}
]

\section{Introduction}

Ni thin films are widely used in spintronic, magnonic and sensing devices, and their magnetic properties are different from those of bulk Ni. Due to the large magnetostriction of Ni, its anisotropy is rather sensitive to strain~\cite{weiler2010acoustically,mathews2023controlling}. Besides, in thin films, surfaces, interfaces and the stress which develops during growth give additional thickness-dependent contributions to the anisotropy~\cite{farle1998ferromagnetic,sander1999correlation,abadias2018stress,jafri2021thickness}. Therefore, the anisotropy of a film has to be determined for the given substrate, growth method and thickness.

Ferromagnetic resonance (FMR) is a standard technique for studying these properties~\cite{farle1998ferromagnetic,gerhardter1993temperature,yalcin2013ferromagnetic,madduri2018ferromagnetic}. The resonance field gives the $g$-factor and the effective magnetization, which combines the demagnetizing field with the perpendicular anisotropy, whereas the linewidth contains the Gilbert damping together with extrinsic contributions from inhomogeneity and two-magnon scattering~\cite{fermin1999ferromagnetic,arias1999extrinsic,lindner2003non}. Often, the resonance is measured only with the field in the film plane and along the normal, and the $g$-factor and the effective magnetization are obtained from the Kittel equations. However, this analysis assumes that the field is exactly aligned with the film and that higher-order anisotropy terms are absent, and otherwise the $g$-factor is biased~\cite{shaw2013precise}. When the resonance is measured as a function of the field direction, the $g$-factor, the anisotropy constants and the exact sample orientation can be determined together~\cite{farle1998ferromagnetic,chappert1986ferromagnetic,zakeri2007spin,masood2024measurement}.

In this work, sputtered Ni films with thicknesses of 300 and 1000\,\AA{} on Si(111) are studied. The structure is characterized by atomic force microscopy, grazing-incidence X-ray diffraction and X-ray reflectivity, and the magnetization by magnetometry. Then, the angular dependence of the FMR from the film plane to the film normal is used to determine the $g$-factor and the first- and second-order perpendicular anisotropies, which are compared with the conventional two-geometry analysis. Finally, an upper limit of the Gilbert damping is obtained from the linewidth. The theoretical background is summarized in Sec.~\ref{sec:theory}, the experimental details are given in Sec.~\ref{sec:exp}, and the results are discussed in Sec.~\ref{sec:results}.

\section{Theoretical background}
\label{sec:theory}

\subsection{Resonance condition}

In an FMR experiment, the motion of the magnetization $\vect{M}$ is described by the Landau--Lifshitz--Gilbert equation~\cite{gilbert2004phenomenological}
\begin{equation}
  \frac{d\vect{M}}{dt} = -\gamma\,\vect{M}\times\vect{H}_{\mathrm{eff}} + \frac{\alpha}{\Ms}\,\vect{M}\times\frac{d\vect{M}}{dt},
  \label{eq:llg}
\end{equation}
where $\vect{H}_{\mathrm{eff}}$ is the effective field, which includes the external field together with the internal demagnetizing and anisotropy fields, $\gamma = g\mu_B/\hbar$ is the gyromagnetic ratio and $\alpha$ is the Gilbert damping parameter. When the damping term is neglected, the resonance frequency $\omega$ is obtained from the free energy density $F$ as~\cite{baselgia1988derivation,suhl1955ferromagnetic,farle1998ferromagnetic}
\begin{equation}
  \begin{split}
  \left(\frac{\omega}{\gamma}\right)^{2} = \frac{1}{\Ms^{2}\sin^{2}\theta}\Biggl[&\frac{\partial^{2}F}{\partial\theta^{2}}\,\frac{\partial^{2}F}{\partial\phi^{2}}\\
  &- \left(\frac{\partial^{2}F}{\partial\theta\,\partial\phi}\right)^{2}\Biggr],
  \end{split}
  \label{eq:sbgen}
\end{equation}
where $\theta$ and $\phi$ are the polar and azimuthal angles of $\vect{M}$. The derivatives are evaluated at the equilibrium angles, which minimize $F$.

For a polycrystalline film with a perpendicular uniaxial anisotropy and without in-plane anisotropy, the free energy density is given by~\cite{farle1998ferromagnetic,johnson1996magnetic}
\begin{equation}
  \begin{split}
  F = {}&-\Ms H\left[\cos\theta\cos\theta_H + \sin\theta\sin\theta_H\cos(\phi - \phi_H)\right]\\
  &+ \left(2\pi\Ms^{2} - K_1\right)\cos^{2}\theta - K_2\cos^{4}\theta .
  \end{split}
  \label{eq:F}
\end{equation}
Here, $\theta_H$ and $\phi_H$ are the polar and azimuthal angles of the external field $\vect{H}$, and all polar angles are measured from the film normal. The first term is the Zeeman energy, the second term is the demagnetizing energy together with the first-order perpendicular anisotropy $K_1$, and the third term is the second-order perpendicular anisotropy $K_2$. Positive values of $K_1$ and $K_2$ favour a magnetization along the normal. For convenience, the effective magnetization and the second-order anisotropy field are defined as
\begin{equation}
  \Meff = 4\pi\Ms - \frac{2K_1}{\Ms}, \qquad H_4 = \frac{4K_2}{\Ms}.
  \label{eq:meff}
\end{equation}
Since the film is isotropic in the plane, the magnetization lies in the plane formed by $\vect{H}$ and the film normal ($\phi = \phi_H$), and its polar angle is obtained from $\partial F/\partial\theta = 0$:
\begin{equation}
  H\sin(\theta - \theta_H) = \frac{\Meff}{2}\sin 2\theta - H_4\cos^{3}\theta\sin\theta .
  \label{eq:equil}
\end{equation}
By inserting Eq.~(\ref{eq:F}) into Eq.~(\ref{eq:sbgen}), the resonance condition is obtained as~\cite{farle1998ferromagnetic,zakeri2007spin}
\begin{equation}
  \left(\frac{\omega}{\gamma}\right)^{2} = H_1H_2,
  \label{eq:sb}
\end{equation}
with
\begin{subequations}
\label{eq:h1h2}
\begin{align}
  H_1 &= H\cos(\theta - \theta_H) - \Meff\cos 2\theta \nonumber\\
      &\quad + H_4\left(\cos^{4}\theta - 3\cos^{2}\theta\sin^{2}\theta\right),\\
  H_2 &= H\cos(\theta - \theta_H) - \Meff\cos^{2}\theta + H_4\cos^{4}\theta ,
\end{align}
\end{subequations}
and $\gamma/2\pi = 1.3996\,g$\,MHz\,Oe$^{-1}$. For the field in the film plane ($\theta_H = 90^\circ$) and along the normal ($\theta_H = 0^\circ$), Eq.~(\ref{eq:sb}) reduces to the Kittel equations~\cite{kittel1948theory,farle1998ferromagnetic}
\begin{subequations}
\label{eq:kittel}
\begin{align}
  \left(\frac{\omega}{\gamma}\right)^{2} &= H_{R\parallel}\left(H_{R\parallel} + \Meff\right), \label{eq:kittelpar}\\
  \frac{\omega}{\gamma} &= H_{R\perp} - \Meff + H_4 . \label{eq:kittelperp}
\end{align}
\end{subequations}
Therefore, the in-plane resonance field depends on $\Meff$, whereas the perpendicular resonance field depends on $\Meff - H_4$. If $H_4$ is neglected, $g$ and $\Meff$ can be determined from the two resonance fields. However, if $H_4 \neq 0$, the full angular dependence is required to separate $g$, $\Meff$ and $H_4$.

Using Eq.~(\ref{eq:meff}), the anisotropy constants are obtained from $\Meff$, $H_4$ and $\Ms$ as
\begin{equation}
  \begin{gathered}
  K_1 = \frac{\Ms}{2}\left(4\pi\Ms - \Meff\right), \qquad K_2 = \frac{\Ms H_4}{4},\\
  \Keff = K_1 + K_2 .
  \end{gathered}
  \label{eq:K}
\end{equation}
According to Eq.~(\ref{eq:F}), the energy of a magnetization along the normal is higher than that of an in-plane magnetization by $2\pi\Ms^{2} - \Keff$. Thus, the film remains magnetized in plane as long as $\Keff < 2\pi\Ms^{2}$.

\subsection{Linewidth}

The peak-to-peak linewidth of the FMR line can be written as the sum of a Gilbert contribution, an inhomogeneous contribution and a two-magnon contribution~\cite{zakeri2007spin,shigemi2001study}:
\begin{equation}
  \Delta H = \Delta H_{\mathrm{G}} + \Delta H_{\mathrm{inh}} + \Delta H_{\mathrm{2M}} .
  \label{eq:lw}
\end{equation}
The Gilbert contribution is given by~\cite{zakeri2007spin}
\begin{equation}
  \Delta H_{\mathrm{G}} = \frac{2}{\sqrt{3}}\,\frac{\alpha\,\omega}{\gamma\cos\beta},
  \label{eq:gilbert}
\end{equation}
where $\beta = \theta - \theta_H$ is the angle between the magnetization and the field. This contribution is proportional to the frequency and becomes larger at intermediate angles, where the magnetization lags behind the field. The inhomogeneous contribution originates from a spread of the local resonance fields, for example due to a distribution of $\Meff$ or of the local film orientation, and it is independent of the frequency~\cite{chappert1986ferromagnetic,zakeri2007spin}. The two-magnon contribution results from the scattering of the uniform mode into degenerate spin waves at defects. It is present for in-plane magnetization, whereas it vanishes when the magnetization is tilted by more than $45^\circ$ out of the film plane~\cite{arias1999extrinsic,zakeri2007spin}. Therefore, for the field along the normal ($\beta = 0$, $\Delta H_{\mathrm{2M}} = 0$), the linewidth is $\Delta H_{\perp} = (2/\sqrt3)\,\alpha\,\omega/\gamma + \Delta H_{\mathrm{inh}}$, and since $\Delta H_{\mathrm{inh}} \ge 0$,
\begin{equation}
  \alpha \le \frac{\sqrt3}{2}\,\frac{\Delta H_{\perp}}{\omega/\gamma}.
  \label{eq:alphaup}
\end{equation}
In order to obtain a value of $\alpha$ rather than an upper limit, the frequency dependence of the linewidth is required, which separates the Gilbert term from the frequency-independent contributions~\cite{shigemi2001study,zakeri2007spin,peria2020interplay}.

\section{Experimental details}
\label{sec:exp}

\subsection{Sample preparation}

Ni films were grown on Si(111) wafers at ambient temperature using a dc magnetron sputtering system~\cite{singh2025growth}. Before the deposition, the base pressure of the chamber was $2\times10^{-6}$\,mbar; the deposition was carried out at an argon sputtering pressure of $4\times10^{-3}$\,mbar with a sputtering power of 40\,W applied to the Ni target. The film thickness was set by the deposition time. Films with nominal thicknesses of 300 and 1000\,\AA{} were prepared; they are denoted \Nia{} and \Nib.

\subsection{Structural and magnetic characterization}

The surface topography was obtained by atomic force microscopy (AFM) in contact mode over $5\times5$\,$\mu$m$^2$ areas. The structure was studied with a high-resolution X-ray diffractometer (HRXRD) using Cu K$\alpha$ radiation ($\lambda = 1.5406$\,\AA), in grazing-incidence X-ray diffraction (GIXRD) mode over $2\theta = 10$--$90^\circ$ at room temperature, and the film thickness by X-ray reflectivity (XRR).

Magnetization measurements were carried out at 300\,K with a SQUID vibrating sample magnetometer (SQUID-VSM), part of the physical property measurement system (PPMS) of Quantum Design Inc., USA, in fields of up to $\pm20$\,kOe applied in the film plane and along the normal. Because the measured moment contains the diamagnetic contribution of the Si substrate, this contribution was subtracted following the procedure of Ref.~\cite{akkera2015martensitic}. The substrate susceptibility $\chi_{\mathrm{sub}} = (dm/dH)_{\mathrm{sub}}$ was determined from the slope of the $M(H)$ loop of a bare Si(111) substrate, and the magnetization of the Ni film was obtained as
\begin{equation}
  M(H) = \frac{1}{V}\left[m_{\mathrm{total}}(H) - \chi_{\mathrm{sub}}\,\frac{m_{\mathrm{sample}}}{m_{\mathrm{sub}}}\,H\right],
  \label{eq:msub}
\end{equation}
where $m_{\mathrm{total}}$ is the measured moment, $m_{\mathrm{sample}}$ and $m_{\mathrm{sub}}$ are the masses of the sample and of the bare substrate, and $V$ is the volume of the Ni film, given by the film area and the XRR thickness. $M$ is expressed in emu\,cm$^{-3}$~\cite{cullity2011introduction}. The saturation magnetization $\Ms$ was taken as the mean $|M|$ at 18--20\,kOe in both loops.

\subsection{Ferromagnetic resonance}

FMR spectra were recorded at 300\,K using a JES-FA200 (JEOL) ESR spectrometer operating at a frequency of $f = 9.443$\,GHz. Because field modulation is used, the spectra represent the field derivative of the absorbed microwave power, $dP/dH$. The sample was rotated about an axis in the film plane, so that the static field turned from the film plane (rotation angle $0^\circ$) through the normal ($90^\circ$) to the plane again ($180^\circ$), in $10^\circ$ steps. The rotation angle is therefore related to the field angle of Eq.~(\ref{eq:F}) by $\theta_H = |90^\circ - \text{rotation angle}|$, apart from a small misalignment of the sample, which is determined from the data.

Above 1.5\,kOe, where all resonances lie, each spectrum was fitted by least squares~\cite{marquardt1963algorithm} with a sum of lines of the form~\cite{dyson1955electron,kalarickal2006ferromagnetic}
\begin{equation}
  \frac{dP}{dH} = A\,\frac{-2\Gamma x\cos\varphi + \left(\Gamma^{2} - x^{2}\right)\sin\varphi}{\left(x^{2} + \Gamma^{2}\right)^{2}},\quad x = H - H_R,
  \label{eq:lineshape}
\end{equation}
on a smooth background. Here, $A$ is the amplitude, $H_R$ the resonance field, $\Gamma$ the half width at half maximum of the absorption, and $\varphi$ the phase that describes an admixture of the dispersive signal, which is common for metallic films~\cite{dyson1955electron,kalarickal2006ferromagnetic}. The peak-to-peak linewidth of each line is $\Delta H = 2\Gamma/\sqrt{3}$~\cite{kalarickal2006ferromagnetic}. Additional lines were accepted only if they improved the fit significantly according to the Bayesian information criterion~\cite{schwarz1978estimating}.

Following earlier angle-dependent FMR studies of thin films~\cite{farle1998ferromagnetic,heinrich1993ultrathin,zakeri2007spin}, Eqs.~(\ref{eq:equil})--(\ref{eq:h1h2}) were then fitted to the resonance fields of the main line at all angles, with $g$, $\Meff$, $H_4$ and the orientation of the film normal as free parameters. For each field direction, the equilibrium angle $\theta$ was obtained from Eq.~(\ref{eq:equil}), and the resonance field was obtained from Eq.~(\ref{eq:sb}). Since $H_R$ changes by up to ${\approx}200$\,Oe per degree near the normal, each point was weighted with an uncertainty which combines 5\,Oe in field with $0.5^\circ$ in angle.

\section{Results and discussion}
\label{sec:results}

\subsection{Structure}
\label{sec:struct}

\begin{figure}[t]
  \centering
  \includegraphics[width=1\columnwidth]{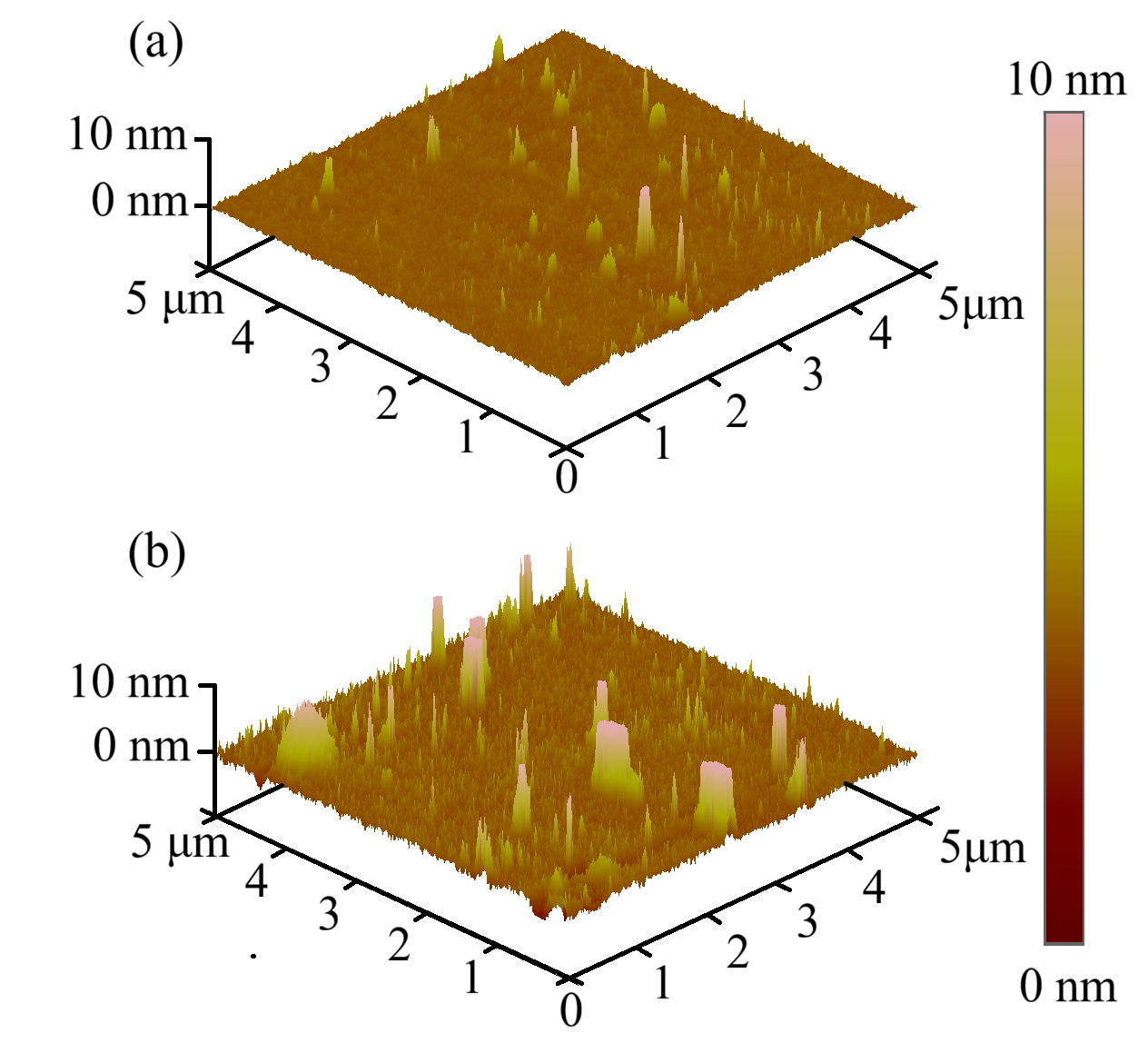}
  \caption{AFM topography ($5\times5$\,$\mu$m$^2$) of (a) \Nia{} and (b) \Nib{} on Si(111).}
  \label{fig:afm}
\end{figure}

The AFM images are shown in Fig.~\ref{fig:afm}. Both surfaces are smooth, with an rms roughness of $\sigma = 4.6$\,\AA{} for \Nia{} and 15.8\,\AA{} for \Nib{}. The higher roughness of the thicker film is mainly due to isolated protrusions with heights of up to ${\sim}10$\,nm. However, the relative roughness, $\sigma/t \approx 1.5\%$, is the same for both films (Table~\ref{tab:struct}).

\begin{figure}[t]
  \centering
  \includegraphics[width=\columnwidth]{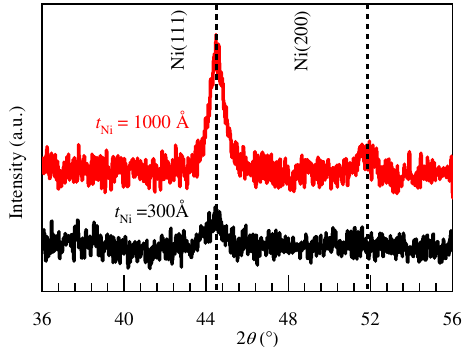}
  \caption{GIXRD patterns of \Nib{} and \Nia; dotted lines mark the Ni(111) and Ni(200) positions of bulk fcc Ni.}
  \label{fig:xrd}
\end{figure}

The GIXRD patterns displayed in Fig.~\ref{fig:xrd} show the reflection of Ni(111) at $2\theta \approx 44.5^\circ$ for both films, and for \Nib{} a weak reflection of Ni(200) is also observed at ${\approx}51.8^\circ$, in agreement with fcc Ni (JCPDS 04-0850~\cite{jcpds}). No reflections of NiO or nickel silicides are detected. The (111) lattice spacings of $2.037\pm0.007$ and $2.033\pm0.007$\,\AA{} are equal to the bulk value of 2.034\,\AA{} within the uncertainty. Therefore, any homogeneous strain along the measured direction is below ${\approx}0.3\%$. The crystallite size $D$ was estimated from the width of the (111) reflection using the Scherrer equation~\cite{SR2026Fe,patterson1939scherrer,vigneshwari2026evolution}
\begin{equation}
  D = \frac{K\lambda}{\beta_{hkl}\cos\theta_{hkl}},
  \label{eq:scherrer}
\end{equation}
where $K = 0.91$ is the shape factor, $\beta_{hkl}$ is the full width at half maximum of the reflection (in radians) and $\theta_{hkl}$ is the Bragg angle. This gives $D = 174$\,\AA{} for \Nia{} and 264\,\AA{} for \Nib. Since Eq.~(\ref{eq:scherrer}) neglects strain broadening, these values are lower limits~\cite{patterson1939scherrer}.

\begin{figure*}[t]
  \centering
  \includegraphics[width=0.9\textwidth]{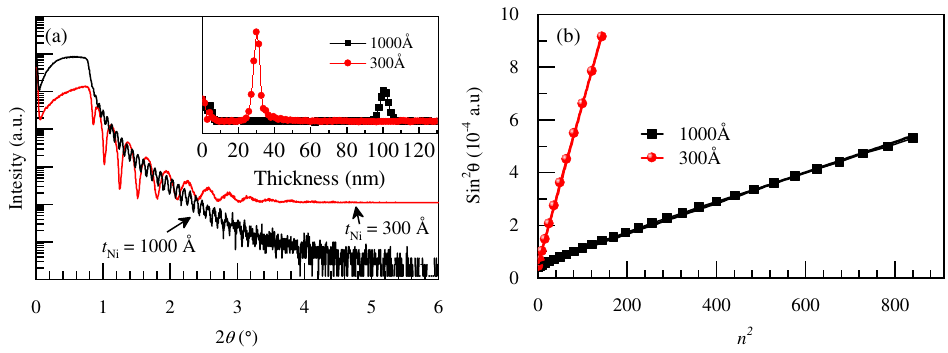}
  \caption{(a) XRR curves of \Nib{} and \Nia; inset: Fourier transform of the reflectivity of both films. (b) $\sin^2\theta_n$ versus $n^2$ for the fringe maxima with linear fits according to Eq.~(\ref{eq:bragg}).}
  \label{fig:xrr}
\end{figure*}

The XRR curves illustrated in Fig.~\ref{fig:xrr}(a) show Kiessig fringes up to $2\theta \approx 3$--$4^\circ$, indicating homogeneous films with well-defined interfaces. The positions of the fringe maxima follow the modified Bragg law~\cite{kiessig1931interferenz,yasaka2010x,wang1993study}
\begin{equation}
  \sin^2\theta_n = \left(\frac{\lambda}{2t}\right)^{2} n^2 + 2\delta,
  \label{eq:bragg}
\end{equation}
where $\theta_n$ is the angle of the $n$th maximum, $t$ is the film thickness, and $\delta$ is the decrement of the real part of the refractive index. Linear fits of $\sin^2\theta_n$ versus $n^2$ presented in Fig.~\ref{fig:xrr}(b) give $t = 312$ and 1011\,\AA, which agree with the nominal thicknesses and with the Fourier transform of the reflectivity [308 and 1010\,\AA, inset of Fig.~\ref{fig:xrr}(a)]. The parameters obtained from the structural analysis are summarized in Table~\ref{tab:struct}.

\begin{table}[t]
  \centering
  \small
  \setlength{\tabcolsep}{4pt}
  \caption{Structural parameters.}
  \label{tab:struct}
  \begin{tabular}{lcc}
    \toprule
    & \Nia & \Nib \\
    \midrule
    Thickness $t$, XRR (\AA) & 312 & 1011 \\
    rms roughness $\sigma$, AFM (\AA) & 4.6 & 15.8 \\
    $d_{111}$ (\AA)$^{a}$ & $2.037\pm0.007$ & $2.033\pm0.007$ \\
    Crystallite size $D$ (\AA) & 174 & 264 \\
    \bottomrule
  \end{tabular}

  \smallskip
  {\footnotesize $^{a}$Bulk Ni: 2.034\,\AA{} (JCPDS 04-0850).}
\end{table}

\subsection{Static magnetization}
\label{sec:static}

\begin{figure}[t]
  \centering
  \includegraphics[width=0.85\columnwidth]{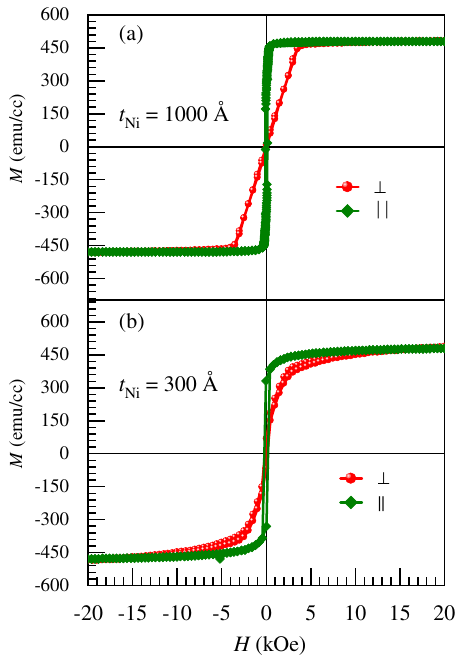}
  \caption{Magnetization loops at 300\,K with the field in the film plane ($\parallel$) and along the normal ($\perp$) for (a) \Nib{} and (b) \Nia.}
  \label{fig:mag}
\end{figure}

The magnetization ($M$--$H$) loops measured at 300\,K are shown in Fig.~\ref{fig:mag} for (a) \Nib{} and (b) \Nia. For both films, the magnetization saturates at much lower fields with the field in the film plane than along the normal, which shows that the films are magnetized in plane. \Nib{} shows a nearly square in-plane loop, whereas along the normal the magnetization increases almost linearly with the field and saturates at ${\approx}4$\,kOe. This field is well below $4\pi\Ms = 6.0$\,kOe, which already indicates a perpendicular anisotropy; this anisotropy is quantified by FMR in Sec.~\ref{sec:aniso}. For \Nia, the magnetization approaches saturation gradually in both directions, which may indicate a distribution of local magnetic anisotropies or a non-uniform magnetization in part of the film~\cite{maity2026octahedral}. At 18--20\,kOe, both films reach a saturation magnetization of $\Ms \approx 480$\,emu\,cm$^{-3}$, close to the bulk value of 484\,emu\,cm$^{-3}$~\cite{cullity2011introduction}.

\begin{figure*}[t]
  \centering
  \includegraphics[width=1\textwidth]{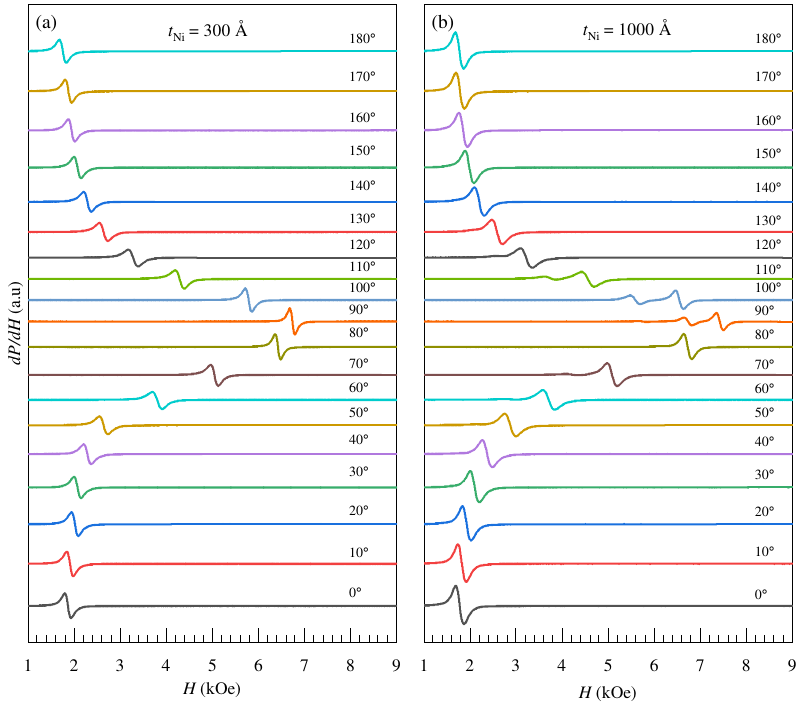}
  \caption{FMR spectra ($dP/dH$) at 9.443\,GHz for (a) \Nia{} and (b) \Nib{} at rotation angles from $0^\circ$ (field in the film plane) through $90^\circ$ (near the normal) to $180^\circ$, in steps of $10^\circ$. The spectra are vertically offset for clarity.}
  \label{fig:fmr}
\end{figure*}

\begin{figure}[t]
  \centering
  \includegraphics[width=0.5\textwidth]{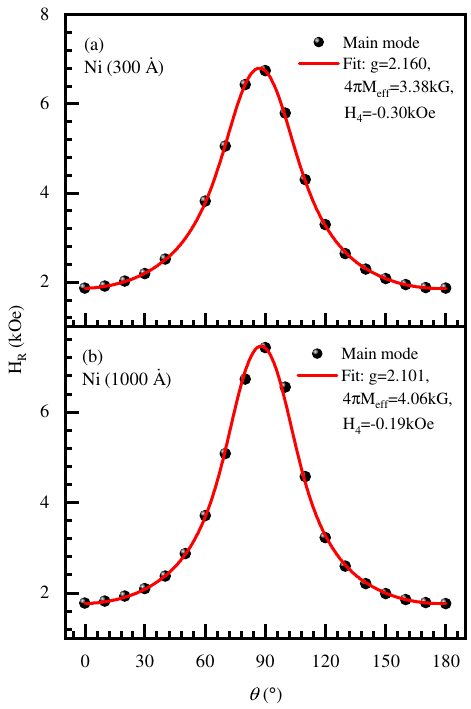}
  \caption{Resonance field of the main line versus rotation angle ($0^\circ$ and $180^\circ$: field in the film plane, $90^\circ$: near the normal) for (a) \Nia{} and (b) \Nib. Symbols: data; red curves: fits of Eqs.~(\ref{eq:equil})--(\ref{eq:h1h2}) with the parameters of Table~\ref{tab:mag}.}
  \label{fig:adfmr}
\end{figure}

\subsection{FMR spectra}
\label{sec:spectra}

Fig.~\ref{fig:fmr} shows the FMR spectra of both films at all rotation angles. With the field in the plane ($0^\circ$ and $180^\circ$), both films show a single line at $H_{R\parallel} \approx 1.8$\,kOe. As the field is rotated towards the normal, the resonance shifts to higher fields and reaches 6.74\,kOe for \Nia{} and 7.42\,kOe for \Nib{} close to the normal (Table~\ref{tab:mag}). The higher perpendicular resonance field of the thicker film corresponds to a larger $\Meff - H_4$ [Eq.~(\ref{eq:kittelperp})].

\Nia{} shows a single line at all angles. On the other hand, close to the normal, \Nib{} shows two weaker resonances below the main line [Fig.~\ref{fig:fmr}(b), Table~\ref{tab:modes}]; the stronger of them is resolved up to ${\approx}38^\circ$ from the normal, whereas in the plane the lines are not resolved. Such resonances may arise from standing spin waves across the film thickness~\cite{kittel1958excitation,qin2018exchange,dkabrowski2021canted} or from regions with a different effective magnetization~\cite{portis1963low,bhoi2013fmr}. Since a definite assignment requires frequency- or thickness-dependent measurements, these lines are not analysed further, and the following analysis uses only the main line.

\begin{table}[t]
  \centering
  \small
  \setlength{\tabcolsep}{4pt}
  \caption{Resonance lines of \Nib{} in the spectrum closest to the normal ($\theta_H = 2.2^\circ$).}
  \label{tab:modes}
  \begin{tabular}{cccc}
    \toprule
    Line & $H_R$ (kOe) & $\Delta H$ (Oe) & Relative intensity$^{a}$ \\
    \midrule
    1 (main) & 7.42 & 142 & 1.00 \\
    2 & 6.74 & 167 & 0.63 \\
    3 & 5.75 & 194 & 0.08 \\
    \bottomrule
  \end{tabular}

  \smallskip
  {\footnotesize $^{a}$Integrated absorption relative to line 1.}
\end{table}

\subsection{Angular dependence: \texorpdfstring{$g$}{g}-factor and anisotropy}
\label{sec:aniso}

The resonance field of the main line at all angles is shown in Fig.~\ref{fig:adfmr}. $H_R$ is lowest in the plane and highest near the normal, as expected for an in-plane magnetized film, and the curves are symmetric about the normal. The fits of Eqs.~(\ref{eq:equil})--(\ref{eq:h1h2}) (red curves) reproduce the data over the full angular range with rms deviations of 28\,Oe (\Nia) and 97\,Oe (\Nib), and the resulting parameters are listed in Table~\ref{tab:mag}. According to the fits, the film normal is tilted by $3.0^\circ$ (\Nia) and $2.2^\circ$ (\Nib) from the nominal position. For \Nib, the largest deviations occur at the two angles adjacent to the normal and correspond to angular errors of ${\approx}1.5^\circ$, so its parameter uncertainties were scaled accordingly. The second-order term $H_4$ differs from zero by more than seven times its uncertainty for \Nia, whereas for \Nib{} it differs by only about twice its uncertainty. Therefore, $H_4$ is only weakly established for \Nib.

\begin{table}[t]
  \centering
  \small
  \setlength{\tabcolsep}{4pt}
  \caption{Magnetic parameters at 300\,K. Upper block: magnetometry and the spectra closest to the film plane and normal; lower block: fit of Eqs.~(\ref{eq:equil})--(\ref{eq:h1h2}) to the angular dependence of $H_R$, and Eq.~(\ref{eq:K}).}
  \label{tab:mag}
  \begin{tabular}{lcc}
    \toprule
    & \Nia & \Nib \\
    \midrule
    $\Ms$ (emu\,cm$^{-3}$) & 481 & 480 \\
    $H_{R\parallel}$ / $H_{R\perp}$ (Oe)$^{a}$ & 1866 / 6738 & 1771 / 7421 \\
    $\Delta H_{\parallel}$ / $\Delta H_{\perp}$ (Oe)$^{a}$ & 137 / 107 & 169 / 142 \\
    $\alpha$ (upper limit, Eq.~(\ref{eq:alphaup})) & 0.030 & 0.038 \\
    \midrule
    $g$ & $2.160\pm0.005$ & $2.101\pm0.014$ \\
    $\Meff$ (kG) & $3.38\pm0.03$ & $4.06\pm0.08$ \\
    $H_4$ (kOe) & $-0.30\pm0.04$ & $-0.19\pm0.10$ \\
    $K_1$, $K_2$ ($10^{5}$\,erg\,cm$^{-3}$) & 6.40, $-0.36$ & 4.75, $-0.23$ \\
    $\Keff$ ($10^{5}$\,erg\,cm$^{-3}$) & $6.0\pm0.4$ & $4.5\pm0.6$ \\
    rms deviation of fit (Oe) & 28 & 97 \\
    \bottomrule
  \end{tabular}

  \smallskip
  {\footnotesize $^{a}$At $\theta_H = 87.0^\circ$ and $3.0^\circ$ (\Nia), $87.8^\circ$ and $2.2^\circ$ (\Nib).}
\end{table}

Using $\Ms$ from magnetometry, the anisotropy constants listed in Table~\ref{tab:mag} were obtained from Eq.~(\ref{eq:K}). The effective perpendicular anisotropy is $\Keff = (6.0\pm0.4)\times10^{5}$\,erg\,cm$^{-3}$ for \Nia{} and $(4.5\pm0.6)\times10^{5}$\,erg\,cm$^{-3}$ for \Nib, and its uncertainty is mainly due to that of $\Ms$. In both films, $\Keff$ is smaller than the shape anisotropy $2\pi\Ms^2$ ($\Keff/2\pi\Ms^2 = 0.42$ and 0.31), which is consistent with the in-plane magnetization observed in Sec.~\ref{sec:static}. For \Nib, the perpendicular resonance gives $\Meff - H_4 = 4.25$\,kG [Eq.~(\ref{eq:kittelperp})], which agrees with the out-of-plane saturation field of ${\approx}4$\,kOe in the magnetization loop [Fig.~\ref{fig:mag}(a)]. Thus, the static and dynamic measurements give a consistent picture.

Perpendicular anisotropy constants of the same order, $(3$--$7)\times10^{5}$\,erg\,cm$^{-3}$, were reported for polycrystalline Ni films of 10--20\,nm on Si(100), where a $1/t$ dependence was attributed to an interface anisotropy~\cite{walowski2008intrinsic}. The larger $\Keff$ of the thinner film in this work is consistent with such an interface contribution. On the other hand, a volume contribution can arise from magnetoelastic coupling to the stress in the film, for example due to growth stress or the thermal-expansion mismatch with the substrate~\cite{sander1999correlation,abadias2018stress,layadi2020effect,mathews2023controlling}. Since only two thicknesses were studied and the stress was not measured, these contributions are not separated here.

The $g$-factors of $2.160\pm0.005$ and $2.101\pm0.014$ are slightly lower than the value of 2.21 which is commonly used for Ni~\cite{walowski2008intrinsic}. When Eqs.~(\ref{eq:kittelpar}) and (\ref{eq:kittelperp}) are solved with $H_4 = 0$ for the spectra closest to the plane and the normal, $g = 2.121$ and 2.079 are obtained, which are 0.04 and 0.02 lower. This shows that neglecting an anisotropy term which is present in the film biases the $g$-factor obtained at a single frequency or over a limited field range~\cite{shaw2013precise}. For the same reason, and since the angular errors of the \Nib{} measurement and a possible small field offset mainly affect $g$, the difference between the two films is not interpreted.

\subsection{Linewidth}
\label{sec:lw}

The linewidth of the main line is 107\,Oe (\Nia) and 142\,Oe (\Nib) near the normal, and 137 and 169\,Oe in the plane (Table~\ref{tab:mag}). Using the perpendicular linewidths and the $g$-factors of Table~\ref{tab:mag}, Eq.~(\ref{eq:alphaup}) gives upper limits of the Gilbert damping of $\alpha \le 0.030$ for \Nia{} and $\alpha \le 0.038$ for \Nib. Since only one frequency was used, the Gilbert and inhomogeneous contributions in Eq.~(\ref{eq:lw}) cannot be separated, and therefore no value of $\alpha$ is reported. In both films, the in-plane linewidth is larger than the perpendicular one by ${\approx}30$\,Oe. Two-magnon scattering, which acts only for in-plane magnetization, is a common reason for such an excess~\cite{arias1999extrinsic,lindner2003non,zakeri2007spin,peria2020interplay}. However, a different inhomogeneous broadening in the two geometries cannot be excluded from the linewidth at a single frequency~\cite{zakeri2007spin}.

\section{Conclusions}

Ni films of 312 and 1011\,\AA{} were sputtered on Si(111) and studied by structural, magnetometric and angle-dependent FMR measurements. Both films are polycrystalline fcc Ni with smooth surfaces, a bulk-like lattice spacing and saturation magnetization, and in-plane magnetization. The angular dependence of the resonance field, analysed with a uniaxial anisotropy model including a second-order term, gives $g = 2.160$ and 2.101 and effective perpendicular anisotropies of $6.0\times10^{5}$ and $4.5\times10^{5}$\,erg\,cm$^{-3}$ for the thinner and thicker film, respectively; the larger value in the thinner film is consistent with an interface contribution. For the thicker film, the anisotropy from FMR agrees with the out-of-plane saturation field of the magnetization loop. The conventional two-geometry analysis gives $g$-factors which are 0.02--0.04 lower, which shows the importance of the full angular dependence. The perpendicular linewidths limit the Gilbert damping to $\alpha \le 0.030$ and 0.038. Measurements at further thicknesses and frequencies would allow the volume and interface anisotropies, the intrinsic damping and the origin of the weaker resonances of the thicker film to be determined.

\section*{Acknowledgments}
The author KRK gratefully acknowledges the financial support received from the Science and Engineering Research Board (SERB), Department of Science and Technology (DST), Government of India, under Grant No.~EEQ/2022/000354. The authors also thank the Central Instrumentation Facility (CIF), IIT Gandhinagar, for providing access to the instrumentation required for material characterisation. Support from the DST-FIST programme under Grant No.~SR/FST/PS-I/2020/147 (C) is also acknowledged for facilitating the magnetization measurements performed using the Physical Property Measurement System (PPMS) at the Cryogen-Free Measurement Facility, IIT Gandhinagar.

\section*{Declaration of competing interest}
The authors declare that they have no known competing financial interests or personal relationships that could have appeared to have influenced the work reported in this study.

\section*{Data availability}
The data supporting the findings of this study are available from the corresponding author upon reasonable request.

\bibliographystyle{iopart-num-doi}
\bibliography{references}

\end{document}